\documentclass[prl,twocolumn,tightenlines,superscriptaddress,a4paper,lengthcheck]{revtex4-1}
\usepackage{amsfonts}
\usepackage{amsmath}
\usepackage{amssymb}
\usepackage{graphicx}
\usepackage{graphicx,amssymb,amsbsy,color}
\usepackage{bm}
\usepackage{bbold}

\begin{document}

\title{Theory of thermalization in an isolated Bose-Einstein condensate}
\author{Che-Hsiu Hsueh}
\affiliation{Department of Physics, National Taiwan Normal University, Taipei 11677, Taiwan}
\author{Chi-Ho Cheng}
\affiliation{Department of Physics, National Changhua University of Education, Changhua 50007, Taiwan}
\author{Tzyy-Leng Horng}
\email{tlhorng@math.fcu.edu.tw}
\email{tlhorng123@gmail.com}
\affiliation{Department of Applied Mathematics, Feng Chia University, Taichung 40724, Taiwan}
\author{Wen-Chin Wu}
\email{wu@ntnu.edu.tw}
\affiliation{Department of Physics, National Taiwan Normal University, Taipei 11677, Taiwan}

\date{\today}

\begin{abstract}
Thermalization in an isolated oscillating Bose-Einstein condensate in a disordered trap is investigated.
We show Shannon entropy in $x$ or $p$ representation is the eligible one to describe the thermalization.
Besides, we show that multiple scattering with the disorder
generates more and more incoherent thermal particles and
condensed and thermal particles act as mutual heat bath that results in the
thermalization of the whole system.
We also demonstrate that Loschmidt's paradox can be resolved in the present system.
\end{abstract}
\maketitle

It is a longstanding question on the microscopic description of the second law of thermodynamics.
One fundamental issue is whether an isolated system can reach thermal equilibrium,
{\rm i.e.}, the state with maximum entropy
\cite{RevModPhys.83.863,Gogolin_2016,10.1038/nature06838,PhysRevLett.108.094102}.
In quantum systems, it is especially
important to see how the reversible microscopic quantum mechanics
conceals the irreversible macroscopic phenomena of thermodynamics.
Fortunately, ultracold quantum gases, pure and controllable,
provide an excellent platform to study the nonequilibrium dynamics
for isolated quantum systems.
In atomic Bose-Einstein condensate (BEC) experiments,
Kinoshita {\em et al.} \cite{Nature.440.900} showed no evidence of thermalization by
pairwise collision, from the Tonks--Girardeau limit to the intermediate coupling regime.
But the dissipative motion of oscillating BEC in a disordered trap, done by
Dries {\em et al.} \cite{PhysRevA.82.033603}, did manifest the thermalization \cite{PhysRevA.98.063613}.
It inspires to investigate what are the eligible microscopic entropies
and how the thermalization is driven in such a system.

In equilibrium statistical mechanics, one primary task is to determine the distribution function.
Owing to the dependence of interaction, exact distribution function of a many-body system is
usually not accessible.
In nonequilibrium statistical mechanics, in contrast,
the most difficult task is to determine the governing factors
that describe the thermalization. In this Letter, for isolated BEC,
we show that the issue is not on whether there is another new law describing the thermalization,
but on how the present microscopic laws can properly describe it, or on
how the system thermalizes based on these laws. It is believed that
similar mechanism can be applied to other isolated quantum systems.

One popular scenario for thermalization
in isolated quantum systems is the eigenstate thermalization hypothesis (ETH), which assumes
the thermalization at individual eigenstates \cite{PhysRevA.43.2046,PhysRevE.50.888}.
However, ETH does not address how the thermal state is reached from an initial
nonequilibrium state. Concerning the thermalization in
an isolated oscillating BEC with random disorder,
we propose and investigate the following mechanism.
Total wave function of the system can be conceptually
separated into the combination of
condensed ($\psi_{0}$) and thermal ($\psi^\prime$) parts,
$\psi=\psi_{0}+\psi^\prime$, and thus the density is $|\psi|^{2}=|\psi_{0}|^{2}+|\psi^\prime|^{2}+2\Re(\psi_{0}^{\ast}\psi^\prime)$.
The oscillating BEC is initially released from a coherent state with a centroid velocity $v_{0}$, which is an out-of-equilibrium state. Owing to the multiple scattering with disorders,
more and more incoherent thermal particles are generated.
Depending on the free energy of system, the number of thermal particles will eventually saturate.
The incoherence or the random phase of thermal particles guarantees that:
$\int\Re(\psi_{0}^{\ast}\psi^\prime)dx=0$ or equivalently,
total number of particles (including both condensed and thermal parts) is conserved.
Thermal particles help transfer mechanical energy into thermodynamical one,
and eventually cause the thermalization of the whole isolated system.
In other words, during the process of thermalization, condensed and thermal particles
act as mutual heat bath and help each other achieve equilibrium.
The mechanism is similar to the one proposed in a many-body theory of Posazhennikova {\em et al.}
\cite{doi:10.1002/andp.201700124,PhysRevLett.116.225304} who considered quantum tunneling
of a binary system with a central barrier.

We pursue by first considering the dynamic entropy
\begin{equation}\label{Shannon}
  S_{Q}\left(t\right)=-k_{\textrm{B}}\sum_{k}\langle\varphi_{Q{k}}|\hat{\rho}(t)
  |\varphi_{Q{k}}\rangle\ln\left[\langle\varphi_{Q{k}}|\hat{\rho}(t)|\varphi_{Q{k}}\rangle\right],
\end{equation}
introduced by Ingarden \cite{Ingarden1976}. Here $|\varphi_{Q{k}}\rangle$ are eigenstates
of an operator $\hat{Q}$ to be chosen, $\hat{\rho}(t)\equiv|\psi(t)\rangle\langle\psi(t)|$ is the density operator
with $|\psi(t)\rangle$ the time-evolution state of the system.
The production rate of entropy follows
\begin{multline}\label{dShannondt}
  \frac{dS_{Q}\left(t\right)}{dt}=-k_{\textrm{B}}\sum_{k}\langle\varphi_{Q{k}}|\dot{\hat{\rho}}
  |\varphi_{Q{k}}\rangle\ln\left[\langle\varphi_{Q{k}}|\hat{\rho}|\varphi_{Q{k}}\rangle\right] \\
  =\frac{i k_{\textrm{B}}}{\hbar}\sum_{k}\langle\varphi_{Q{k}}|[\hat{H},\hat{\rho}]|\varphi_{Q{k}}
  \rangle\ln[\langle\varphi_{Q{k}}| \hat{\rho}|\varphi_{Q{k}}\rangle],
\end{multline}
where $\dot{\hat{\rho}}\equiv \partial \hat{\rho}/\partial t$ and
$\hat{H}$ is the Hamiltonian. If $\hat{Q}$ is chosen to be either $\hat{H}$ or $\hat{\rho}$,
Eq.~(\ref{dShannondt}) vanishes or $S_Q$ is conserved.
This is the reason why the von Neumann entropy
is conserved and is not eligible to describe the time direction of an isolated quantum system
\cite{PhysRevE.100.012101,POLKOVNIKOV2011486}.

For a one-dimensional (1D) system under consideration,
if $\hat{Q}$ is chosen to be the momentum operator $\hat{p}$ instead, the corresponding entropy is
\begin{equation}\label{entropyp}
  S_{p}\left(t\right)=-k_{\textrm{B}}\int\rho_{kk}\left(t\right)
  \ln\left[\frac{\rho_{kk}\left(t\right)}{a}\right]dk
\end{equation}
and the entropy production rate is
\begin{equation}
\label{dsp/dt}
  \frac{dS_{p}\left(t\right)}{dt}=-k_{\textrm{B}}\int
  \dot{\rho}_{kk}(t)\ln\left[\frac{\rho_{kk}\left(t\right)}{a}\right]dk,
\end{equation}
where $a$ is a characteristic length scale of the system and
$\rho_{kk}\equiv|\widetilde{\psi}(k,t)|^{2}$ is the momentum density distribution
with $\widetilde{\psi}(k,t)=\langle k|\psi(t)\rangle$.
The entropy in Eq.~(\ref{entropyp}) is known as the Shannon entropy in $p$ representation.
Except for a noninteracting system in free space,
$\hat{H}=\hat{p}^2/2m$ and thus $[\hat{H},\hat{p}]=0$, the
entropy $S_p$ in (\ref{entropyp}) is generally not conserved.
Alternatively, if $\hat{Q}$ is chosen to be the position operator $\hat{x}$,
the corresponding entropy
\begin{equation}\label{entropyx}
  S_{x}\left(t\right)=-k_{\textrm{B}}\int\rho\left(t\right)
  \ln\left[a\rho\left(t\right)\right]dx,
\end{equation}
where $\rho(t)\equiv|{\psi}(x,t)|^{2}$ is the density distribution in real space
with ${\psi}(x,t)=\langle x|\psi(t)\rangle$. According to the commutation relation
$[\hat{x},\hat{p}]=i\hbar\neq 0$, one immediately sees that $S_{x}$ in (\ref{entropyx})
is not conserved even for a noninteracting system in free space.

We propose that Shannon entropy in $p$ representation, $S_{p}$ in (\ref{entropyp}) or
in $x$ representation, $S_{x}$ in (\ref{entropyx}), is the proper one to
describe the losing information inherent in an isolated quantum system.
On an equal footing, the sum of the two, $S\equiv S_{x}+S_{p}$, which takes into
account all possible phase space, should be another good candidate.
As a matter of fact, owing to the commutation relation $[\hat{x},\hat{p}]=i\hbar\neq 0$,
$dS_{x}/dt$ and $dS_{p}/dt$ won't vanish simultaneously except for an equilibrium state.

\begin{figure}[tb]
\begin{center}
\includegraphics[height=2.8in,width=3.5in]{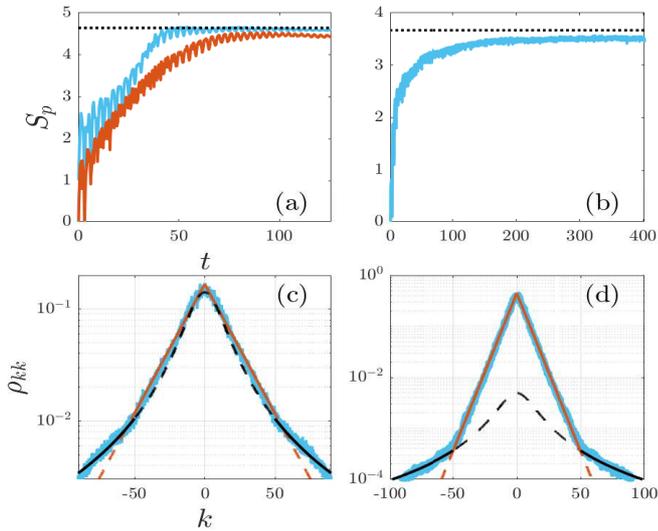}
\end{center}
\vspace{-0.6cm}
\caption{Evolutions of kinetic entropy $S_p(t)$ for the oscillating disordered condensate
with initial velocity $v_{0}=3.75c_{0}$ (a) and $1.4c_{0}$ (b) (blue curves).
Red curve in (a) is the result for a LHY fluid (see the text).
Black dotted lines correspond to the microcanonical $S_{\rm M}$ in (\ref{Boltzmann_entropy}).
(c) \& (d): Momentum distributions $\rho_{kk}$, corresponding to equilibrium states of (a) and (b),
reveal distinct behaviors at low and high $k$'s.
Red and black curves are microcanonical and Rayleigh-Jeans fittings to the
low- and high-$k$ parts.}
\label{fig1}
\end{figure}

It is intended to investigate the thermalization in isolated quantum systems
from the point of view of the total wavefunction $\psi$.
For the isolated oscillating BEC in a disordered trap,
the most comprehensive and simple Hamiltonian is given by the
Gross-Pitaevskii equation (GPE). One can easily go
beyond the mean-field level to include the Lee-Huang-Yang (LHY) quantum corrections
with the local density approximation, $g_{\rm LHY}|\psi|^3$
\cite{PhysRevLett.117.205301,PhysRevLett.119.215302,PhysRevLett.121.195301}.
However, the basic mechanism leading to thermalization remains the same
[see Fig.~\ref{fig1}(a)].

Since the entropy production rate $dS_{x}/dt$
cannot give much direct information on the cause of entropy production,
for illustration purpose, we focus on $dS_{p}/dt$.
In momentum space, GPE reads as
\begin{eqnarray}\label{GPk}
 && i\hbar\partial_{t}\widetilde{\psi}(k,t)=\frac{\hbar^{2}k^{2}}{2m}\widetilde{\psi}(k,t) \nonumber\\
 &&+\frac{1}{2\pi}\int\widetilde{V}\left(k-k_{1}\right)\widetilde{\psi}(k_{1},t)dk_{1} \nonumber\\
  &&+\frac{g}{4\pi^{2}}\int\widetilde{\psi}^{\ast}(k_{3},t)
  \widetilde{\psi}(k_{2},t)\widetilde{\psi}(k_{1},t)\delta_{12}^{3k}dk_{1}dk_{2}dk_{3},
\end{eqnarray}
where $g$ is particle interaction coupling,
$\widetilde{\psi}(k,t)$ and $\widetilde{V}(k)$ are respectively Fourier transforms of
wave function $\psi(x,t)$ and external potential $V(x)$,
and $\delta_{12}^{3k}\equiv\delta\left(k_{1}+k_{2}-k_{3}-k\right)$.
It is important to retain the information of external potential and particle interaction.
Following (\ref{GPk}), after some derivations we obtain $\rho_{kk}(t)$
obeys the following master equation:
\begin{multline}\label{masterequation}
  \dot{\rho}_{kk}=-\frac{1}{\pi\hbar}\int\Im\left[\widetilde{V}^{\ast}\left(k-k_{1}\right)\rho_{kk_{1}}\right]dk_{1} \\
  -\frac{g}{2\pi^{2}\hbar}\int\Im\left[\int\rho_{\left(k_{1}+k_{2}-k\right)k_{2}}dk_{2}\rho_{kk_{1}}\right]dk_{1},
\end{multline}
where $\rho_{kq}\equiv\widetilde{\psi}(k,t)\widetilde{\psi}^{\ast}(q,t)$ and
$\Im[A]$ denotes the imaginary part of $A$.
The first (second) line corresponds to the effect of external potential (particle interaction).

Substitution of (\ref{masterequation}) into (\ref{dsp/dt}) gives
\begin{eqnarray}\label{htheorem}
  &&\frac{dS_{p}\left(t\right)}{dt}=\frac{k_{\textrm{B}}}{2\pi\hbar}\int\Im\left[\widetilde{V}^{\ast}
  \left(k-k_{1}\right)\rho_{kk_{1}}\right]\ln\left(\frac{\rho_{kk}}{\rho_{k_{1}k_{1}}}\right)dk_{1}dk \nonumber\\
  &&+\frac{gk_{\textrm{B}}}{4\pi^{2}\hbar}\int\Im\left[\int\rho_{\left(k_{1}+k_{2}-k\right)k_{2}}dk_{2}
  \rho_{kk_{1}}\right]\ln\left(\frac{\rho_{kk}}{\rho_{k_{1}k_{1}}}\right)dk_{1}dk.\nonumber\\
\end{eqnarray}
It clearly shows that there are two main causes for entropy production:
external potential and particle interaction.
Entropy saturates when the system reaches detailed balancing,
$\rho_{kq}=0$ for $k\neq q$ or equipartition,
$\ln\left(\rho_{kk}/\rho_{k_{1}k_{1}}\right)=\ln\left(1\right)=0$.
For the latter of equal probability, the entropy reaches the maximum,
$S_Q=k_{\textrm{B}}\ln\Omega_{Q}$ with $\Omega_{Q}$ the total number of eigenstates $|\varphi_{Q{k}}\rangle$
and the production rate (\ref{htheorem}) vanishes.
Eq.~(\ref{htheorem}) can be regarded as the second law (or H-theorem) of thermodynamics
for isolated BEC.

External potentials can be classified according to the effect on
entropy production. Without loss of generality, we consider a confining potential
$V(x)=x^{2\alpha}$ with $\alpha$ a positive integer,
of which the Fourier transformation is $\widetilde{V}(k)=i^{2\alpha}2\pi\delta^{(2\alpha)}(k)$.
The delta function will result the first integral in Eq.~(\ref{htheorem}) vanish. It implies
the confining potential makes no contribution to the entropy growth.
This is in agreement with the previous experiment \cite{Nature.440.900} which confirmed that there is
no evidence of redistribution of momentum due to a confining trap or pairwise collisions.
More generally, potentials belonging to the $L^{1}$-nonintegrable function produce no entropy.
In contrast, $L^{1}$-integrable potentials, even a single Gaussian barrier,
may produce entropy \cite{PhysRevA.82.033603}.

In addition to the parabolic trapping potential,
a random Gaussian correlated disorder potential
$V_{\textrm{dis}}(x)$ is added to the oscillating BEC,
which leads to the dissipative
motion (localization) of the condensate \cite{PhysRevA.82.033603}.
Experimentally the condensate gains an initial velocity $v_0$ by
abrupt displacement of the potential \cite{PhysRevA.82.033603}.
Due to multiple scattering with the random disorder,
the system will eventually reach equilibrium from nonequilibrium.
Theoretically initial oscillating condensate can be described by a coherent wavefunction
$\psi(x,0)\simeq\psi_{\textrm{g}}(x)\exp\left(imv_{0}x/\hbar\right)$,
where $\psi_{\textrm{g}}(x)$ is the ground state \cite{PhysRevA.98.063613}.
Consequently the initial momentum distribution
$\rho_{kk}(0)\simeq|\widetilde{\psi}_{\textrm{g}}(k-mv_{0}/\hbar)|^{2}$, which is just a $k$-shift
from the one for the ground state $|\widetilde{\psi}_{\textrm{g}}(k)|^{2}$.
It implies free energy of the oscillating condensate is
higher than that of the ground state, but the
initial entropy $S(0)$ is the same (minimum) as that of the ground state.

Entropy evolutions $S_p(t)$ of the oscillating BEC in random disorder
with initial velocity $v_0=3.75c_0$ and $1.4c_0$ ($c_0$ is sound speed)
are numerically solved and shown in Fig.~\ref{fig1}(a) \& \ref{fig1}(b).
One key feature is that $S_p(t)$ increases and saturates
at $t=t_{\rm th}$, after which the thermalization is achieved. The thermalization time
$t_{\rm th}\simeq 40$ and $200$ in case (a) and (b), respectively.
The black dotted lines correspond to the microcanonical $S_{\rm M}$ given in (\ref{Boltzmann_entropy}).
To see the effect of LHY quantum correction, in Fig.~\ref{fig1}(a) we also plot the
case (red curve) for $g_{\rm LHY}=16000$ and $g=0$, so-called the LHY fluid \cite{PhysRevLett.121.173403}.
The LHY fluid is parameterized to have the same ``Thomas-Fermi radius" as the one without
the LHY correction.
It suggests that the basic mechanism leading to thermalization remains the same.
In our simulation, $\hbar\omega$ ($\omega$ is the trapping frequency) and the characteristic
trapping length $a=\sqrt{\hbar/m\omega}$
are taken as the units of energy and length.
Following the experiment \cite{PhysRevA.82.033603}, the parameters are
$\omega=2\pi\times5.5$Hz,
$\mu=200\hbar\omega$ which gives the sound speed $c_{0}\simeq 10a\omega$, and $g=16000/3$.
For random disorder, the strength $V_{\textrm{D}}=560/11\hbar\omega$ and
the correlation length $\sigma_{\textrm{D}}=0.01a$.

When the system reaches equilibrium ($t > t_{\rm th}$), momentum distributions $\rho_{kk}$
are shown in Fig.~\ref{fig1}(c) \& \ref{fig1}(d) for the case (a) and (b). It is
seen that equilibrium distributions in both cases
exhibit very distinct behaviors at low- and high-$k$ regimes.
At low $k$'s, momentum distribution is
well fitted by the ``microcanonical" distribution (red curve):
\begin{equation}\label{ensemble}
  \rho_{\textrm{M}}=\frac{1}{\Omega_{\textrm{M}}}\exp\left(-\sqrt{\frac{2\beta\hbar^{2}k^{2}}{m}}\right),
\end{equation}
where $\Omega_{\textrm{M}}=\sqrt{2m/(\beta\hbar^{2})}$ is the partition function
and $\beta=1/k_{\textrm{B}}T$.
The equilibrium temperature can be defined as $T_{\rm eq}\equiv2K\left(t\rightarrow\infty\right)/k_{\textrm{B}}$
with the kinetic energy $K(t)=\int|\hbar\partial_{x}\psi|^{2}dx/(2m)$.
Using Eq.~(\ref{ensemble}), ensemble average of the kinetic energy gives $\int \left(\hbar^{2}k^{2}/2m\right)\rho_{\textrm{M}}dk=k_{\textrm{B}}T/2$. The two
temperatures are found to be consistent, $T=T_{\rm eq}$, which
supports the ergodic hypothesis.
In the limit $\beta\rightarrow\infty$, $\rho_{\textrm{M}}\rightarrow
\delta\left(k\right)$ corresponding to the momentum distribution for the
equilibrium ground-state condensate at $T\rightarrow 0$.
It strongly suggests that Eq.~(\ref{ensemble}) corresponds to the distribution
of condensed particles at $T$.
Boltzmann entropy of the distribution (\ref{ensemble}) is
\begin{equation}\label{Boltzmann_entropy}
  S_{\textrm{M}}=k_{\textrm{B}}\ln\left(\mathrm{e}a\Omega_{\textrm{M}}\right).
\end{equation}
As shown in Fig.~\ref{fig1}(a) \& (b), $S_{\textrm{M}}$ matches
the maximum kinetic entropy.

At high $k$'s, the momentum distributions is instead
best fitted by the Rayleigh-Jeans (RJ) distribution:
\begin{equation}\label{RJ}
  \rho_{\textrm{RJ}}=\frac{1}{\Omega_{\textrm{RJ}}}\frac{\hbar\omega}{\left(\hbar^{2}k^{2}/m^*\right)+\mu},
\end{equation}
where $\Omega_{\textrm{RJ}}=\sqrt{\pi^{2}m^*\omega^{2}/\mu}$ with $m^*$ the effective mass
and $\mu$ the chemical potential.
The RJ distribution (\ref{RJ}) resembles the distribution of classical particles.
Thus it unambiguously corresponds to thermal particles.
The picture of two distinct distributions fully supports
the proposed mechanism for thermalization.


\begin{figure}[tb]
\begin{center}
\includegraphics[height=2.0in,width=3.5in]{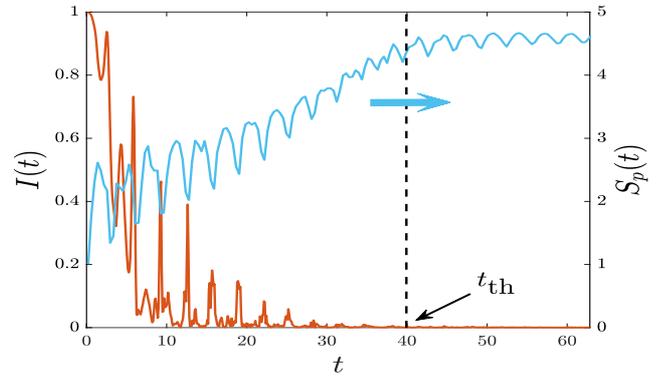}
\end{center}
\vspace{-0.6cm}
\caption{Loschmidt echo as a function of time (see the text). It oscillates but
diminishes initially and vanishes eventually at $t=t_{\rm th}$ when thermalization is reached.}
\label{fig2}
\end{figure}

To further study the correlation between condensed and thermal particles, we
consider the Loschmidt echo function:
\begin{equation}
m(t)\equiv \langle\psi|e^{i\hat{H}t/\hbar} e^{-i\hat{H}_0t/\hbar}|\psi\rangle,
\label{Lc}
\end{equation}
where $\hat{H}=\hat{H}_0+\hat{V}_{\rm dis}$ with $\hat{V}_{\rm dis}$
the random disorder potential. We note that
the state with no effect of disorder, $\psi_0(t)=e^{-i\hat{H}_0t/\hbar}|\psi\rangle$, is a
coherent oscillatory state with a constant centroid velocity $v_0$. Loschmidt echo measures how
the real state (with disorder) overlaps with $\psi_0(t)$ as a function of time.
Fig.~\ref{fig2} shows $I(t)\equiv |m(t)|^2$ for the case shown in Fig.~\ref{fig1}(a).
For a easy comparison, $S_p(t)$ is also shown. As depicted in the current model,
owing to multiple scattering with disorder,
more and more incoherent thermal particles are generated until saturation.
This mechanism is clearly shown in Loschmidt echo in
which $I(t)$ oscillates but diminishes initially and eventually vanishes at $t=t_{\rm th}$.
The incoherence or the random phase of thermal particles causes the vanishing of $I(t)$.

Finally we discuss the time reversibility of entropy production rate (\ref{dsp/dt}).
It is well known that all microscopic laws are time reversible.
It means the reversed evolution of system follows the same physics law as the one of forward evolution,
{\rm i.e.}, there would be no new physics law resulted from time reversal.
Defining the time-reversal operator: $\widehat{T}f\left(x,k,t\right)=f^{\ast}(x,k^\prime,t^\prime)$
with $k^\prime=-k$ and $t'\equiv-t$, and applying it on
Eq.~(\ref{dsp/dt}), one obtains
\begin{equation}\label{dsp/dt'}
  \frac{dS_{p}(t^\prime)}{dt^\prime}=-k_{\textrm{B}}\int \dot{\rho}_{k^\prime k^\prime}(t^\prime)\ln\left[\frac{\rho_{k^\prime k^\prime}(t^\prime)}{a}\right]dk^\prime,
\end{equation}
which shares the same form as Eq.~(\ref{dsp/dt}). Thus the entropy production rate is also time reversible.
Loschmidt's paradox says that if the time direction is reversed,
the entropy of isolated system will decrease, and it will violate the thermodynamic second law.
This statement is only partially correct because the entropy cannot decrease without a limit.
Time-reversed process will eventually hits the origin,
which might be one of the lowest-entropy states, and one may quest what happens subsequently?
Because of the symmetry in both Eqs.~(\ref{dsp/dt'}) and (\ref{dsp/dt}),
the entropy will also increase \emph{before} the origin.

\begin{figure}[tb]
\includegraphics[height=2.3in,width=3.3in]{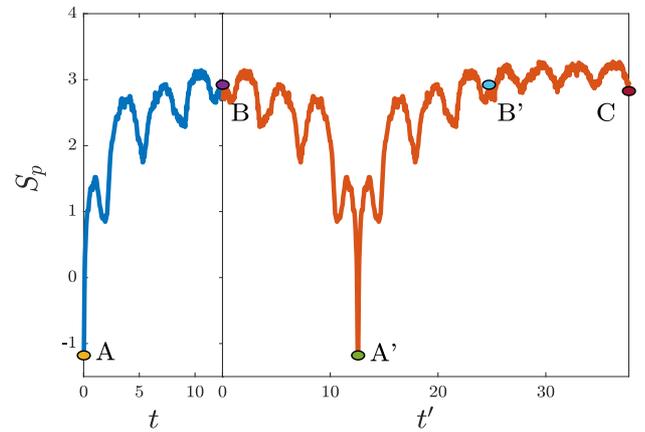}
\caption{Entropy evolution of an oscillating BEC ($v_{0}=1.4c_{0}$)
in a harmonic potential with a Gaussian barrier at the center.
Forward evolution (blue) from state A to B ($t=0$ to $4\pi$);
inverse evolution (red) from B to C ($t^\prime =0$ to $12\pi$).
State A$^\prime$ (at $t^\prime=4\pi$) indicates a revival to A and corresponds to
a mirror point to which entropy increases regardless time flows forward or backward.}
\label{fig3}
\end{figure}

Here we give a real example how Loschmidt's paradox is resolved.
Fig.~\ref{fig3} shows the evolution of kinetic entropy (\ref{entropyp}) for
an oscillating condensate (with $v_{0}=1.4c_{0}$)
in a harmonic potential with a Gaussian barrier at the center.
Blue curve shows the forward evolution from state A ($t=0$) to B ($t=4\pi$),
after that the evolution is reversed to along the inverse direction of time,
shown as the red curve, from state B ($t^\prime=0$) to C ($t^\prime=12\pi$).
The state A$^\prime$($t^\prime=4\pi$) indicates it revives to the initial one at A,
and after that, the entropy follows the same process to increase (from A$^\prime$ to B$^\prime$),
exactly as the one from A to B.
It implies that there exists a mirror symmetry at A$^\prime$ with the minimum entropy, and
at this point, the system tends to maximize the entropy regardless time flows forward or backward.
Such a revival phenomenon is universal.

Recently a new cosmology model suggests that our universe has a mirror image in the form of an antiuniverse
that existed before the big bang \cite{PhysRevLett.121.251301}.
This model is capable to answer several unsolved questions, {\em e.g.},
why does the initial state of the universe have an extraordinarily low entropy,
why does the universe mainly consist of matter, not antimatter,
and where goes the primitive antimatter generated from the big bang?
The initial state (A or A$^\prime$) shown in Fig.~\ref{fig3} can be viewed as
the preferred \emph{CPT}--invariant state at the big bang,
which is coherent and thus has an extraordinarily low entropy. From
such an initial state, the universe constructed of matters follows the positive-time evolution,
while the antiuniverse constructed of antimatters follows the negative-time direction, and
entropies of both universe and antiuniverse {\em increase}.

In conclusion, a theory is proposed for the thermalization in an
isolated oscillating BEC with random disorder. It is aimed to study the
phenomena exhibited in the total wavefunction $\psi(t)$.
We clearly show that during the process, generation of more and more
incoherent thermal particles results in the thermalization.
We also demonstrate the Loschmidt's paradox
can be resolved in the present system.
Moreover, the system can be linked to a newly proposed cosmology model that
considers a preferred \emph{CPT}--invariant state at the big bang \cite{PhysRevLett.121.251301}.

Financial support from the Ministry of Science and Technology, Taiwan
(under grant No. 108-2112-M-003-006-MY2) is acknowledged.


%

\end{document}